\documentclass[epj-spec]{svjour}
\usepackage{graphicx}
\begin{document}
\title{Spectroscopy of atomic hydrogen}
\subtitle{How is the Rydberg constant determined?}
\author{F. Biraben\inst{1}\fnmsep\thanks{\email{biraben@spectro.jussieu.fr}}}
\institute{Laboratoire Kastler Brossel, ENS, CNRS, UPMC, 4 place Jussieu, 75252 Paris CEDEX 05, France}
\abstract{
This article presents a review of the most recent theoretical and experimental results in hydrogen. We particularly emphasize the methods used to deduce the Rydberg constant $R_\infty$ and we consider the prospects for future improvements in the precision of $R_\infty$.
} %end of abstract
\maketitle
\section{Introduction}
\label{intro}
\begin{figure}
\centering
%\resizebox{0.75\columnwidth}{!}{%
\includegraphics[width=10cm]{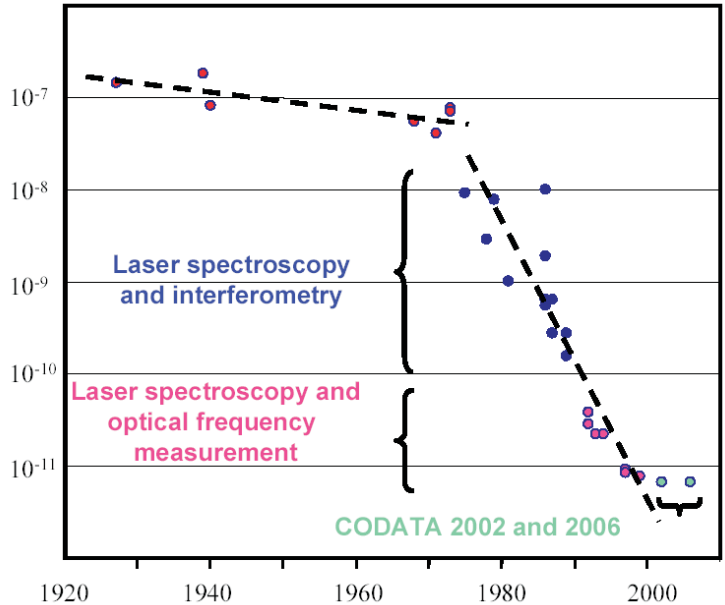}
\caption{Relative precision of the Rydberg constant from 1920 to the present.}
\label{fig:Rydberg}       % Give a unique label
\end{figure}

The hydrogen atom has a central position in the history of 20th-century physics. As it is the simplest of atoms, it has played a key role in testing fundamental theories, and hydrogen spectroscopy is associated with successive advances in the understanding of atomic structure. The advent of tunable lasers and nonlinear techniques of Doppler free spectroscopy in the seventies led to major advances in resolution and measurement precision which are described in references \cite{{SciAmerican},{Series}}. Since, in the nineties, hydrogen spectroscopy has inspired a revolution in the art of measuring the frequency of light thanks to optical frequency combs \cite{Udem}. Consequently, several optical frequencies of hydrogen are now known with a fractional accuracy better than $10^{-11}$, the most precise being the 1S-2S two photon transition which has been measured with a relative uncertainty of $1.4\times10^{-14}$ \cite{1S-2Sa}. Thanks to these advances, the accuracy of the Rydberg constant $R_\infty$ has been improved by several orders of magnitude in three decades. This is illustrated in the figure \ref{fig:Rydberg} which clearly shows the improvements due to laser spectroscopy and optical frequency measurements. Nevertheless, during the last decade, there has been little progress with, for example, no important improvement between the $R_\infty$ values  given by the last two adjustments of the fundamental constants in 2002 and 2006 \cite{{codata02},{codata06}}. In this review, we describe the analysis of the theoretical and experimental data used to deduce $R_\infty$ and we discuss the prospects for the future improvements in the accuracy of $R_\infty$.

\section{Energy levels of atomic hydrogen: theoretical calculations}
\label{sec:2}
\subsection{Theoretical background}
\label{sec:2-1}
\begin{figure}
\centering
%\resizebox{0.75\columnwidth}{!}{%
\includegraphics[width=8cm]{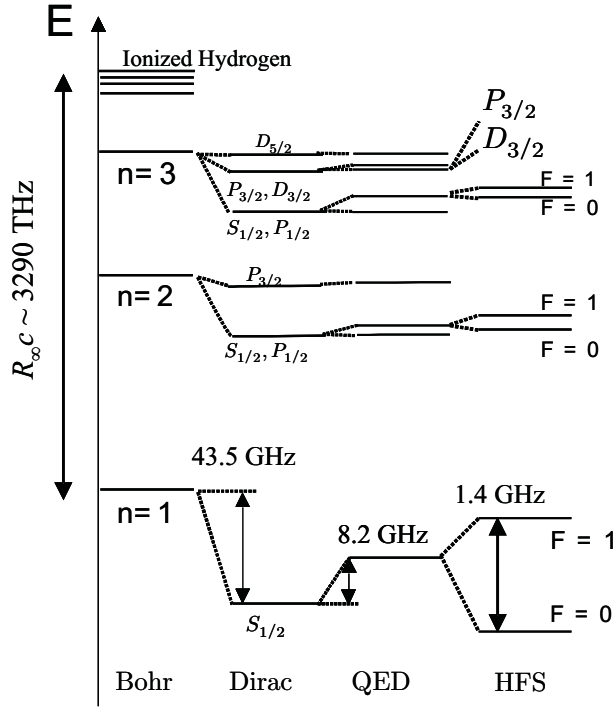}
\caption{Energy levels of atomic hydrogen for successive steps of the theory. The Dirac equation describes the fine structure splitting between the $j=1/2$ and $j=3/2$ levels and QED theory the splitting between the $2S_{1/2}$ and $2P_{1/2}$ levels.}
\label{fig:Niveaux}       % Give a unique label
\end{figure}

Figure \ref{fig:Niveaux} shows the energy levels of hydrogen for different steps of the theory. The solution of the Schr\"{o}dinger equation gives the same energy levels as the simple Bohr model. The energy $E_n$ depends only on the principal quantum number $n$:
\begin{equation}
E_n=-\frac{hcR_\infty}{n^2}
\label{equ:Bohr}
\end{equation}
The Rydberg constant $R_\infty$ is known as a function of the fine structure constant $\alpha$, of the velocity of light $c$, of the Planck constant $h$ and of the electron mass $m_e$:
\begin{equation}
R_\infty=\frac{{\alpha}^2m_ec}{2h}
\label{equ:Rydberg}
\end{equation}
This equation which links several fundamental constants is very useful for the adjustment of the fundamental constants. In particular, as $R_\infty$ is very well known, it is a link between the fine structure constant $\alpha$ and the $h/m_e$ ratio and it is a way to deduce $\alpha$ from the $h/m$ measurements \cite{{Kruger},{Wicht},{Clade1},{Clade2}}.

The next step takes into account the relativistic corrections which are given by the Dirac equation. This equation lifts the degeneracy in $j$ ($j$ is the total angular momentum: $j=l\pm1/2$) and explains the fine structure, but, for instance, the levels $2S_{1/2}$ and $2P_{1/2}$ are degenerate. This degeneracy disappears with the corrections due to quantum electrodynamics (QED),which are responsible for the Lamb shift between the $2S_{1/2}$ and $2P_{1/2}$ levels, first observed by Lamb and Retherford in 1947 \cite{Lamb}. In this paper, we wish to give only an idea of the theoretical calculations in hydrogen and more details are found in text books and recent review papers \cite{{Bethe},{Eides},{Karshenboim1}}.

The hydrogen level energy can be conventionally expressed by the sum of three terms:
\begin{equation}
E_{n,l,j}=E_{n,j}^{\mathrm{Dirac}}+E_{n}^{\mathrm{Recoil}}+L_{n,l,j}
\label{equ:Energie}
\end{equation}
where $E_{n,j}^{\mathrm{Dirac}}$ is the energy given by the Dirac equation for a particle with a reduced mass $m_r=m_e(1+m_e/m_p)^{-1}$ and $E_{n}^{\mathrm{Recoil}}$ the first relativistic correction due to the recoil of the proton which depends only on the principal quantum number $n$. The last term $L_{n,l,j}$ is the Lamb shift which takes into account all the other corrections: the QED corrections, the other relativistic corrections due to the recoil of the proton and the effect of the proton charge distribution. The first two terms of equation \ref{equ:Energie} are exactly known as a function of the Rydberg constant $R_\infty$, of the fine structure constant $\alpha$ and of the electron to proton mass ratio $m_e/m_p$. In contrast, the calculation of the Lamb shift $L_{n,l,j}$ is very difficult. It is obtained in a series of terms in powers of $\alpha$, $Z\alpha$ ($Z$ is the charge of the nucleus and $Z\alpha$ characterizes the interaction between the proton and the electron), $m_e/m_p$ and the root-mean-square charge radius of the proton $r_p$. A very clear review of these different terms is given in the reference \cite{codata02}. In this paper, we only present the most recent results.

\subsection{Recoil corrections}
\label{sec:2-2}
The first recoil correction is:
\begin{equation}
E_{n}^{\mathrm{Recoil}}=-\frac{m_r^2c^2}{m_e+m_p}\frac{(Z\alpha)^4}{8n^4}
\label{equ:Recul}
\end{equation}
In terms of frequency, this first correction is about 24 MHz for the 1S level, i.e. $10^{-8}$ times the ionization energy of hydrogen. The following terms of the recoil correction vary as $m_ec^2(m_e/m_p)^2(Z\alpha)^4$ and $m_ec^2(m_e/m_p)(Z\alpha)^5$ and they have an exact expression. The next term, which varies as $m_ec^2(m_e/m_p)(Z\alpha)^6$, has been calculated in the period 1988-1998 by several authors \cite{{Erickson-Grotch1},{Erickson-Grotch2},{Erickson-Grotch3},{Fell},{Pachucki},{Shabaev}}. Table \ref{tab:1} summarizes the successive results which have been obtained and shows the difficulty of this kind of calculations. Finally, the value of this correction is now known with an uncertainty of 10 Hz for the 1S level.

\begin{table}
\caption{Successive calculations of the term in $m_ec^2(m_e/m_p)(Z\alpha)^6$ of the correction due to the recoil of the proton.}
\label{tab:1}       % Give a unique label
\begin{tabular}{lll}
\hline\noalign{\smallskip}
Reference & Mathematical expression & Value for the 1S level (MHz)  \\
\noalign{\smallskip}\hline\noalign{\smallskip}
\cite{Erickson-Grotch1}&$m_ec^2(m_e/m_p)(Z\alpha)^6n^{-3}[3-\ln(2/Z\alpha)+...]$ &$-26.6~\rm kHz$\\
\cite{Erickson-Grotch2}&$m_ec^2(m_e/m_p)(Z\alpha)^6n^{-3}[5/2-\ln(2/Z\alpha)+...]$& $-31.8~\rm kHz$\\
\cite{Erickson-Grotch3}&$m_ec^2(m_e/m_p)(Z\alpha)^6n^{-3}[5/2-\ln(2/Z\alpha)+2\ln(1/Z\alpha)-4.25]$ & $+25.2~\rm kHz$\\
\cite{Fell}&There is no term in $\ln(1/Z\alpha)$ &  \\
\cite{Pachucki}&$m_ec^2(m_e/m_p)(Z\alpha)^6n^{-3}[4\ln(2)-7/2]$ & $-7.4~\rm kHz$\\
\cite{Shabaev}&Numerical calculation to all orders in $Z\alpha$& $-7.16(1)~\rm kHz$ \\
\hline\noalign{\smallskip}
\end{tabular}
\end{table}
\subsection{Quantum electrodynamics corrections}
\label{sec:2-3}
The corrections due to QED are the main contribution to the Lamb shift. The self energy (SE) corresponds to the emission and reabsorption of virtual photons by the electron, and the vacuum polarization (VP) to the creation of virtual electron-positron pairs. A simple explanation of the self energy is given by the Welton model \cite{Welton}. Because of the residual energy of the empty modes of the electromagnetic field (the energy $\hbar\omega/2$ of the harmonic oscillators), the electron is submitted to the fluctuations of the vacuum field which induce fluctuations in its position. This effect modifies the Coulomb potential seen by the electron and is particularly important for the S levels: it reduces the binding energy, i.e. it increases slightly the energy of the S states ($l=0$), because, for the S states, the electron has a large probability $|\Psi(0)|^2$ to be inside the nucleus. This is the reason for the splitting between the $2S_{1/2}$ and $2P_{1/2}$ levels (see Figure \ref{fig:Niveaux}).

For the self energy, the lowest-order radiative correction (called "one-loop" or "one-photon" correction) is given by:
\begin{equation}
E_\mathrm{SE}^{(2)}=\frac{\alpha}{\pi}\frac{(Z\alpha)^4}{n^3}F(Z\alpha)m_ec^2
\label{equ:SE1}
\end{equation}
where $F(Z\alpha)$ is a sum of terms in powers of $Z\alpha$ and $\ln(Z\alpha)$:
\begin{equation}
F(Z\alpha)=A_{41}\ln(Z\alpha)^{-2}+A_{40}+A_{50}(Z\alpha)+A_{62}(Z\alpha)^2\ln^2(Z\alpha)^{-2}+A_{61}(Z\alpha)^2\ln(Z\alpha)^{-2}+G_\mathrm{SE}(Z\alpha)(Z\alpha)^2
\label{equ:FSE1}
\end{equation}
There are similar equations for the vacuum polarization correction. All the terms in equation \ref{equ:FSE1} have an exact expression, except the last one $G_\mathrm{SE}(Z\alpha)$ which has been calculated intensively since the seventies \cite{{Mohr1},{Erickson},{Sapirstein81},{Pachucki1},{Jentschura},{Jentschura1}}. More recently, the one-loop correction to all orders in $Z\alpha$ has been obtained numerically for the $1S$, $2S$, $3S$ and $4S$ levels with an uncertainty of a few Hz \cite{{Jentschura2},{Jentschura3}}.

The following radiative correction takes into account the emission and reabsorption of two virtual photons. It varies as $\alpha^2$ and is given by:
\begin{equation}
E^{(4)}=\left(\frac{\alpha}{\pi}\right)^2\frac{(Z\alpha)^4}{n^3}F^{(4)}(Z\alpha)m_ec^2
\label{equ:SE2}
\end{equation}
where:
\begin{equation}
F^{(4)}(Z\alpha)=B_{40}+B_{50}(Z\alpha)+B_{63}(Z\alpha)^2\ln^3(Z\alpha)^{-2}+B_{62}(Z\alpha)^2\ln^2(Z\alpha)^{-2}+B_{61}(Z\alpha)^2\ln(Z\alpha)^{-2}+B_{60}(Z\alpha)^2+...
\label{equ:FSE2}
\end{equation}
The terms of equation \ref{equ:FSE2} are calculated in references \cite{{Pachucki2},{Eides1},{Pachucki3},{Eides3},{Mallampalli}}. However there is a controversy on the last calculated term $B_{60}$. For the 1S level, Pachucki and Jentschura obtain $B_{60}=-61.6(9.2)$ \cite{Pachucki4} while Yerokin et al. give $B_{60}=-127(38)$ \cite{Yerokin}. This difference corresponds to 6.6 kHz for the Lamb shift of the 1S level.

\subsection{Other corrections}
\label{sec:2-4}
There are many of other terms in the calculation of the Lamb shift which are detailed in reference \cite{codata02}: the three-loop radiative corrections, the effect of the creation of virtual pairs $\mu^+\mu^-$ and $\tau^+\tau^-$, the radiative recoil corrections, the self energy and the polarization of the nucleus, and the effect of the non-zero size of the nucleus. This last term is important and is given for the proton by:
\begin{equation}
E_\mathrm{NS}=\frac{2}{3}\left(\frac{m_r}{m_e}\right)^3\frac{(Z\alpha)^2}{n^3}m_ec^2\left(\frac{2\pi Z\alpha r_p}{\lambda_C}\right)^2
\label{equ:PS}
\end{equation}
where $\lambda_C$ is the Compton wavelength of the electron. The rms charge radius of the proton $r_p$ is obtained from elastic electron-proton scattering experiments. There is a long history of analysis of these experiments. The most recent work gives $r_p=0.895(18)~\mathrm{fm}$ which corresponds to a shift of the 1S level  of about $1.2~\mathrm{MHz}$ \cite{Sick}.

\begin{table}
\caption{Summary of the calculation of the 1s Lamb shift. The main uncertainties are due to the proton size correction (about $50~\rm kHz$) and to the two-loop corrections (about $3.3~\rm kHz$). The uncertainties in the one-loop correction (SE and VP) are essentially due to the uncertainty in the fine structure constant $\alpha$.}
\label{tab:2}       % Give a unique label
\begin{tabular}{lrr}
\hline\noalign{\smallskip}
Term of the Lamb shift & Value for the 1S level & Uncertainties \\
\noalign{\smallskip}\hline\noalign{\smallskip}
Self-energy (one-loop)&$8\,383\,339.466~\rm kHz$ &$0.083~\rm kHz$\\
Vacuum polarization (one-loop)&$~-214\,816.607~\rm kHz$&$0.005~\rm kHz$ \\
Recoil corrections &$~~~2\,401.782~\rm kHz$&$0.010~\rm kHz$ \\
Proton size &$~~~1\,253.000~\rm kHz$&$50~\rm kHz$ \\
Two-loop corrections &$~~~~731.000~\rm kHz$&$3.300~\rm kHz$ \\
Radiative recoil corrections &$~~~~-12.321~\rm kHz$&$0.740~\rm kHz$ \\
Vacuum polarization (muon)&$~~~~~-5.068~\rm kHz$&$<0.001~\rm kHz$ \\
Vacuum polarization (hadron)&$~~~~~-3.401~\rm kHz$&$0.076~\rm kHz$ \\
Proton self-energy &$~~~~~~4.618~\rm kHz$&$0.160~\rm kHz$ \\
Three-loop corrections &$~~~~~~1.800~\rm kHz$&$1.000~\rm kHz$ \\
Nuclear size corrections to SE and VP &$~~~~~~-0.149~\rm kHz$&$0.011~\rm kHz$ \\
Proton polarization &$~~~~~~~-0.070~\rm kHz$&$0.013~\rm kHz$ \\
\hline\noalign{\smallskip}
1S Lamb shift &$8\,172\,894(51)~\rm kHz$ \\
\hline\noalign{\smallskip}
\end{tabular}
\end{table}
To summarize, we give in table \ref{tab:2} the value of the different terms contributing to the Lamb shift of the 1S level. For this calculation, we have used the value of $r_p$ from reference \cite{Sick} and the values of $R_\infty$, $\alpha$ and $m_e/m_p$ given in reference \cite{codata02}. The theoretical uncertainty in the 1S Lamb shift is estimated to be $3.7~\rm kHz$, mainly due to the calculation of the two-loop corrections. At this level, the precision is not limited by the uncertainties in $R_\infty$, $\alpha$ and $m_e/m_p$. On the contrary, the uncertainty due to the charge distribution of the proton $r_p$ is about $50~\rm kHz$, making it the largest source of uncertainty in the calculation of the 1S Lamb shift.

\section{Precise measurements in hydrogen}
\label{sec:3}
In this section, we present the hydrogen frequency measurements which are used for the adjustment of the fundamental constants and for the determination of the Rydberg constant.

\subsection{Lamb shift of the $2S_{1/2}$ level}
\label{sec:3-1}
\begin{figure}
\centering
\includegraphics[width=10cm]{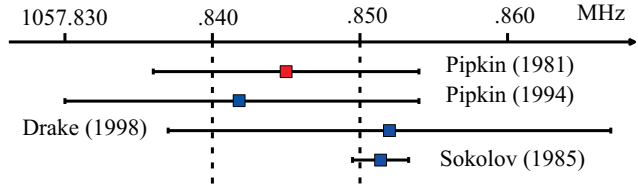}
\caption{Recent measurements of the Lamb shift between the $2\rm S_{1/2}$ and $2\rm P_{1/2}$ levels.}
\label{fig:Lambshift}       % Give a unique label
\end{figure}

Since the historic measurement of Lamb and Retherford \cite{Lamb}, a number of measurements of the $2\rm S_{1/2}-2\rm P_{1/2}$ splitting have been reported \cite{{Newton},{Pipkin1},{Pipkin2},{Drake},{Sokolov}}. The  recent results are shown in the figure \ref{fig:Lambshift}. The most precise direct determination of this splitting is the one by Lundeen and Pipkin ($1057.845(9)~\rm MHz$ \cite{Pipkin1}). The value obtained by Hagley and Pipkin in 1994 is an indirect determination deduced from the $2\rm S_{1/2}-2\rm P_{3/2}$ splitting \cite{Pipkin2}. By using the theoretical value of the $2\rm P_{1/2}-2\rm P_{3/2}$ fine structure splitting \cite{Jentschura}, this determination is $1057.842(12)~\rm MHz$ (see figure \ref{fig:Lambshift}). The result of Drake in 1998 ($1057.852(15)~\rm MHz$ \cite{Drake}) is also an indirect determination, obtained by measuring the anisotropy of the emitted light in an applied electric field. The experiment of Pal'chikov, Sokolov and Yakovlev was performed by using atomic interferometry to measure the ratio between the $2\rm S_{1/2}$ Lamb shift and the lifetime of the $2\rm P_{1/2}$ level. This result is very precise ($1057.8514(19)~\rm MHz$ \cite{Sokolov}), but there is a controversy over the theoretical value of $2\rm P_{1/2}$ lifetime which is used \cite{{Sokolov1},{Karshenboim2}}. Finally, if we take into account only the direct measurements of the $2\rm S_{1/2}-2\rm P_{1/2}$ and $2\rm S_{1/2}-2\rm P_{3/2}$ splittings, we obtain a mean value of $1057.8439(72)~\rm MHz$ for the $2\rm S_{1/2}$ Lamb shift.

\subsection{The 1S-2S transition}
\label{sec:3-2}

The $1\rm S-2\rm S$ transition is studied by Doppler-free two-photon spectroscopy, first proposed by Vasilenko et al. \cite{Chebotayev1}. The principle is to place an atom in a standing wave formed by two counter-propagating laser beams with the same frequency. If the atom absorbs one photon from each beam, the total momentum of the absorbed photons is zero, and, consequently, there is a cancellation of the first order Doppler effect and of the recoil effect. The idea to apply this two-photon spectroscopy to the 1S-2S transition in hydrogen was immediately proposed by several authors, Cagnac et al. \cite{JPhys73}, Baklanov and Chebotayev \cite{Chebotayev2} and H\"{a}nsch et al. \cite{Hansch74}. It was a very attractive proposal: because of the very small natural width ($1.3~\rm Hz$) of the $2\rm S$ level, the quality factor of the $1\rm S-2\rm S$ transition is about $2\times 10^{15}$. However it was experimentally very difficult, due to the UV wavelength of this transition ($243~\rm nm$) and the low transition probability.

Since the first observation by H\"{a}nsch et al. in 1975 \cite{Hansch75}, the $1\rm S-2\rm S$ transition has been studied by several other groups in Southampton \cite{Ferguson86}, Oxford \cite{{Stacey},{Stacey90},{Stacey92}}, Yale \cite{Boshier95} and MIT \cite{Kleppner96}. Nevertheless, the most important work has been performed by the group of H\"{a}nsch, who has continuously studied this transition since these first observations. In a long series of experiments, H\"{a}nsch has improved the precision on the measurement of the $1\rm S-2\rm S$ frequency. The first determination of this frequency used an interferometric method with a calibrated absorption line of $^{130}\mathrm{Te}_2$ \cite{{Ferguson86},{Stacey},{Hansch86},{Hansch89}}. In these experiments, the
accuracy was limited by that of the $^{130}\mathrm{Te}_2$ reference (about $2.7\times 10^{-10}$).  In the nineties, this limitation was overcome thanks to optical frequency measurements. In a first experiment, H\"{a}nsch used a frequency chain which linked the 1S-2S frequency (about $2466~\rm THz$) to a transportable $\mathrm{CH}_4$-stabilized He-Ne frequency standard at $88~\rm THz$ \cite{{Hansch92},{Hansch97}}. Now, this complex frequency chain has been superseded by a femtosecond laser frequency comb, which links in one fell swoop the Cs clock at $9~\mathrm{GHz}$ to the optical frequency. Thanks to this technique, pioneered by Hall and H\"{a}nsch, H\"{a}nsch's group has recently succeeded in measuring the 1S-2S interval with respect to a transportable Cs atomic fountain clock from the SYRTE (formerly Laboratoire Primaire du Temps et des
Fr\'{e}quences) in Paris \cite{{Hansch2000},{1S-2Sa}}. This last measurement reduces the uncertainty to $34~\rm Hz$ ({\it i.e.} a relative uncertainty of about $1.4\times 10^{-14}$). The value obtained for the $1\rm S-2\rm S$ frequency is:
\begin{equation}
\nu_{\mathrm{1S-2S}}=2~466~061~413~187~074~(34)~\mathrm{Hz}
\label{equ:nu1S-2S}
\end{equation}
This result is one of the most precise optical frequency measurements.

\subsection{Two-photon spectroscopy of the $2\mathrm{S}-n\mathrm{S}$ and $2\mathrm{S}-n\mathrm{D}$ transitions}
\label{sec:3-3}

In Paris we began to study the $2\mathrm{S}-n\mathrm{S}$ and $2\mathrm{S}-n\mathrm{D}$ transitions in 1983. These experiments are complementary to those of the $1\mathrm{S}-2\mathrm{S}$ measurements, because the Lamb shift of the $2\mathrm{S}$ level has been measured precisely (see section \ref{sec:3-1}), and, consequently, it is easy to extract the Rydberg constant from the $2\mathrm{S}-n\mathrm{S/D}$ interval. The two-photon $2\mathrm{S}-n\mathrm{S/D}$ transitions are induced in a metastable $2\mathrm{S}$ atomic beam of hydrogen or deuterium collinear with the counter-propagating laser beams. The excitation wavelength is in the near infrared, for example $778~\rm nm$ for the $2\mathrm{S}-8\mathrm{S/D}$ transitions. The details of this experiment are described in references \cite{{Garreau1},{Garreau2},{Garreau3}}.

In our first experiments, we used an interferometric method to compare the hydrogen wavelengths to an iodine stabilized He-Ne laser. With this method, we determined the frequencies of the $2\mathrm{S}-n\mathrm{D}$ transitions in hydrogen and deuterium for the levels $n=8$, 10 and 12 \cite{{Ryd86},{Ryd89}}. The relative accuracy was limited to $1.7\times10^{-10}$ by the standard laser. In 1993, we measured the optical frequencies of the $2\rm S_{1/2}-8\rm S_{1/2}$, $2\rm S_{1/2}-8\rm D_{3/2}$ and $2\rm S_{1/2}-8\rm D_{5/2}$ transitions in hydrogen with a frequency chain using two standard lasers (iodine stabilized and methane stabilized helium-neon lasers) and obtained a precision in the range of $10^{-11}$ \cite{{Ryd92},{Ryd93}}. More recently, we remade these measurements in hydrogen and deuterium with an accuracy better than one part in $10^{11}$ \cite{Ryd97} using a new frequency chain with a new standard laser, namely a diode laser at 778~nm stabilized on the 5S-5D two-photon transition of rubidium. The frequency of this standard was measured with a frequency chain at the (LPTF) \cite{Mes Rubidium97}. Finally, we completed these results by the measurement of the $2\mathrm{S}-12\mathrm{S/D}$ transitions \cite{Ryd99}. A complete report of these experiments is given in reference \cite{EPJD00}. For example, we obtain for the frequency of the $2\rm S_{1/2}-8\rm D_{5/2}$ frequency in hydrogen:
\begin{equation}
\nu_{2\rm S_{1/2}-8\rm D_{5/2}}=770~649~561~581.1~(5.9)~\mathrm{kHz}
\label{equ:nu2S-8D}
\end{equation}
and we have a similar result in deuterium. The relative uncertainty is limited to $7.6\times 10^{-12}$ because of the natural width of the $8\rm D$ level ($572~\mathrm{kHz}$) and the inhomogenous light shift experienced
by the atoms passing through the Gaussian profile of the laser beams.

\subsection{Frequency comparison between hydrogen frequencies}
\label{sec:3-4}

The goal of this method is to avoid absolute frequency measurements. The idea, proposed by H\"{a}nsch, is to compare the $1\rm S-2\rm S$ frequency with transitions whose energies are approximatively one-fourth that of the $1\rm S-2\rm S$ transition (see equation \ref{equ:Bohr}), such as the $2\rm S-4\rm P$ transition \cite{{Hansch75},{Hansch80},{Stacey92},{Boshier95}}, or the $2\rm S-4\mathrm{S/D}$ two-photon transitions \cite{{Hansch92b},{Hansch94},{Hansch95}}. In our group, we have also performed an experiment based on the same idea, but this time by comparing the $1\rm S-3\rm S$ and the $2\rm S-6\mathrm{S/D}$ frequencies \cite{Bourzeix}. The three last experiments \cite{{Boshier95},{Hansch95},{Bourzeix}} provided a determination of the 1S Lamb shift with an uncertainty of about $50~\rm kHz$ and these results are always used in the adjustment of the fundamental constants \cite{{codata02},{codata06}}.

\section{Determination of the Rydberg constant}
\label{sec:4}

The Rydberg constant is deduced from the data described in section \ref{sec:3} through a least squares adjustment. It is possible to make this adjustment with only the hydrogen data, the values of the fine structure constant $\alpha$ and the electron-to-proton mass ratio $m_e/m_p$ being given a priori \cite{EPJD00}, or to perform a global adjustment with the data concerning all the fundamental constants. Since 1998, the CODATA (Committee on Data for Science and Technology) has used the latter method to determine the value of $R_{\infty}$. The value obtained in the 2006 CODATA adjustment is \cite{codata06}:
\begin{equation}
R_{\infty}=10~973~731.568~527(73)~\mathrm{m^{-1}}
\label{equ:Rydberg}
\end{equation}
with a relative uncertainty of $6.6\times10^{-12}$. The advantage of this method is to give the most accurate value, but the drawback is a mixing of all the experimental and theoretical results and it is difficult to see the most important input data. In the following sections, we describe several simple ways to determine the Rydberg constant.

\subsection{Determination of $R_{\infty}$ from the $1\rm S-2\rm S$ interval}
\label{sec:4-1}

In equation \ref{equ:Energie}, the Dirac energy $E_{n,j}^{\mathrm{Dirac}}$ and the recoil energy $E_{n}^{\mathrm{Recoil}}$ are exactly known as a function of the Rydberg constant. Thus, it is possible to rewrite this equation in the form:
\begin{equation}
E_{n,l,j}=a_{n,j}hcR_{\infty}+L_{n,l,j}
\label{equ:Energie1}
\end{equation}
where $a_{n,j}$ is a numerical coefficient which is an exactly known function of $\alpha$ and $m_e/m_p$, and whose value is approximatively given by the Bohr model $ a_{n,j}\approx-1/n^2$ (see equation \ref{equ:Bohr}). Then, the $1\rm S-2\rm S$ frequency is:
\begin{equation}
\nu_{\mathrm{1S-2S}}=(a_{2,1/2}-a_{1,1/2})cR_{\infty}+(L_\mathrm{2S_\mathrm{1/2}}-L_\mathrm{1S_\mathrm{1/2}})/h\approx(3/4)cR_{\infty}+(L_\mathrm{2S_\mathrm{1/2}}-L_\mathrm{1S_\mathrm{1/2}})/h
\label{equ:Energie2}
\end{equation}
where $L_\mathrm{1S_\mathrm{1/2}}$ and $L_\mathrm{2S_\mathrm{1/2}}$ are the Lamb shifts of the $1\rm S$ and $2\rm S$ levels. In this equation, the frequency $\nu_{\mathrm{1S-2S}}$ is known with an uncertainty of $34~\mathrm{Hz}$, but the Lamb shift difference $L_\mathrm{2S_\mathrm{1/2}}-L_\mathrm{1S_\mathrm{1/2}}$ is calculated with a precision of only about $44~\mathrm{kHz}$, because of the uncertainty in the proton radius $r_p$. Consequently, in spite of the very high accuracy of the $1\rm S-2\rm S$ measurement ($1.4\times 10^{-14}$), the Rydberg constant can be deduced from equation \ref{equ:Energie2} with a relative uncertainty of only  ($1.8\times 10^{-11}$).

\subsection{Determination of $R_{\infty}$ from the $2\rm S-8\rm D$ interval}
\label{sec:4-2}

To deduce $R_{\infty}$ from the $2\rm S-8\rm D$ measurement, we can follow the same method. We have:
\begin{equation}
\nu_{2\rm S_{1/2}-8\rm D_{5/2}}=(a_{8,5/2}-a_{2,1/2})cR_{\infty}+(L_\mathrm{8D_{5/2}}-L_{2S_\mathrm{1/2}})/h\approx\left(\frac{1}{4}-\frac{1}{64}\right)cR_{\infty}+(L_\mathrm{8D_{5/2}}-L_\mathrm{2S_{1/2}})/h
\label{equ:Energie2S-8D}
\end{equation}
In this expression, the uncertainty on the theoretical values of the Lamb shifts are respectively  $2.5~\mathrm{Hz}$ and $6.4~\mathrm{kHz}$ for the $8\mathrm{D}_{5/2}$ and $2\mathrm{S}_{1/2}$ levels (because of the $1/n^3$ scaling law of the Lamb shift, the uncertainty on the $2S_{1/2}$ Lamb shift is one eighth that of the $1S_{1/2}$). Taking into account the experimental uncertainty ($5.9~\mathrm{kHz}$, see equation \ref{equ:nu2S-8D}), we can finally extract $R_{\infty}$ from equation \ref{equ:Energie2S-8D} with a relative uncertainty of about ($1.1\times 10^{-11}$).

Another way to obtain $R_{\infty}$ is to combine the $2\rm S_{1/2}-8\rm D_{5/2}$ frequency with the measurement of the $2\rm S_{1/2}$ Lamb shift (see section \ref{sec:3-1}). In this way, it is possible to eliminate the $2\rm S_{1/2}$ level. Indeed, we have:
\begin{equation}
\nu_{2\rm P_{1/2}-2\rm S_{1/2}}+\nu_{2\rm S_{1/2}-8\rm D_{5/2}}=(a_{8,5/2}-a_{2,1/2})cR_{\infty}+(L_{8\rm D_{5/2}}-L_{2\rm P_{1/2}})/h
\label{equ:Energie2S-8Dbis}
\end{equation}
In this case, the theoretical uncertainties on the $2\rm P_{1/2}$ and $8\rm D_{5/2}$ Lamb shifts are very small ($80$ and $2.5~\mathrm{Hz}$) \cite{Jentschura}. With this method, the accuracy of $R_{\infty}$ is limited by the uncertainties in the measurements of the frequencies $\nu_{2\rm P_{1/2}-2\rm S_{1/2}}$ and $\nu_{2\rm S_{1/2}-8\rm D_{5/2}}$. The relative accuracy of $R_{\infty}$ is finally $1.2\times 10^{-11}$. This method is slightly less precise, but it does not use the experimental value for the rms charge radius of the proton.

\subsection{Comparison of the $1\rm S-2\rm S$ and $2\rm S-8\rm D$ intervals}
\label{sec:4-3}

In the two preceding sections, we have seen that the accuracy of the Rydberg constant is limited by the theoretical or experimental uncertainties in the Lamb shifts. In fact, it is possible to avoid this difficulty by using the $1/n^3$ scaling law for the Lamb shift. Numerous terms of the Lamb shift vary with the principal quantum number exactly as $1/n^{3}$ (for instance the effect of the charge distribution of the nucleus), and the deviation from this scaling law has been calculated precisely by Karshenboim \cite{Karshenboim3}. He obtains for the $1\rm S$ and $2\rm S$ Lamb shifts:
\begin{equation}
\Delta_2=(L_{1\rm S_{1/2}}-8L_{2\rm S_{1/2}})/h=-187.232\,(5)~\mathrm{MHz}
\label{equ:L1-8L2}
\end{equation}
A more recent calculation of $\Delta_2$ is given in reference \cite{Pachucki5}. If we combine this equation with equations \ref{equ:Energie2} and \ref{equ:Energie2S-8D}, we obtain a system of 3 equations where the unknowns are $R_{\infty}$, $L_{1\rm S_{1/2}}$ and $L_{2\rm S_{1/2}}$. We can then form a linear combination to eliminate the $1\rm S$ and $2\rm S$ Lamb shifts:
\begin{equation}
7\nu_{2\rm S_{1/2}-8\rm D_{5/2}}-\nu_{1\rm S_{1/2}-2\rm S_{1/2}}\approx\left(\frac{57}{64}\right)cR_{\infty}+7L_{8\rm D_{5/2}}/h+\Delta_2
\label{equ:7nu28-nu12}
\end{equation}
In this expression, the main uncertainty is due to the measurement of the $2\rm S_{1/2}-8\rm D_{5/2}$ frequency ($7\times5.9~\mathrm{kHz}$) and the Rydberg constant is deduced with a relative uncertainty of $1.4\times 10^{-11}$. The advantage of this method is to deduce $R_{\infty}$ without the measurements of the $2\rm S_{1/2}$ Lamb shift and of the proton radius $r_p$. Moreover, this technique is applicable to both hydrogen and deuterium (there is no precise measurement of the $2\rm S_{1/2}$ Lamb shift in deuterium). For instance, from the measurements of the $1\rm S_{1/2}-2\rm S_{1/2}$, $2\rm S_{1/2}-8\rm D_{5/2}$ and $2\rm S_{1/2}-12\rm D_{5/2}$ frequencies in hydrogen and deuterium, one obtains a value of $R_{\infty}$ with a relative uncertainty better than $10^{-11}$ \cite{EPJD00}:
\begin{equation}
R_{\infty}=10~973~731.568~54(10)~\mathrm{m^{-1}}
\label{equ:Rydberg}
\end{equation}
Morover, we can deduce the Lamb shift of the $1\rm S_{1/2}$ and $2\rm S_{1/2}$ levels in hydrogen:
\begin{equation}
L_{1\rm S_{1/2}}/h=8~172.837(26)~\mathrm{MHz}
\label{equ:L1S}
\end{equation}
\begin{equation}
(L_{2\rm S_{1/2}}-L_{2\rm P_{1/2}})/h=1~057.8447(34)~\mathrm{MHz}
\label{equ:L2S}
\end{equation}
In this last result, we have used the theoretical value of the $2\rm P_{1/2}$ Lamb shift ($-12.835\,99\,(8)~\mathrm{MHz}$ \cite{Jentschura}). This value of the $2\rm S_{1/2}$ Lamb shift deduced from the optical frequency measurements is more precise than the direct determinations made by microwave spectroscopy.

Finally, if we take into account the theoretical calculations of the Lamb shift (see section \ref{sec:2}), we can deduce a value of the rms charge radius of the proton ($r_p=0.8765\,(80)~\mathrm{fm}$) which is more precise than the one deduced from the electron-proton scattering experiments. For this calculation, we have used $B_{60}=-94.3$, which is the mean value of the results of the references \cite{{Pachucki4},{Yerokin}} (see section \ref{sec:2-3}).

\section{Conclusion}
\label{sec:5}

In conclusion, we have presented several methods to determine the Rydberg constant. There are several limitations, mainly the uncertainty in the rms charge radius of the proton and the uncertainties in the measurements of the $2\mathrm{S}-n\mathrm{S/D}$ frequencies in hydrogen. To reduce the first limitation, a precise measurement of $r_p$ is ongoing at the Paul Scherrer Institute by spectroscopy of muonic hydrogen. The principle is to measure the $2\rm S-2\rm P$ energy difference in $\mu^-\rm p$ by infrared spectroscopy \cite{Antognini}. In muonic hydrogen, the muon is very close to the proton because its mass is about 207 times that of the electron. Consequently, the effect of the proton charge distribution is about $0.93~\mathrm{THz}$ for the $2\rm S$ level of muonic hydrogen whereas it is only $146~\mathrm{kHz}$ in hydrogen. The details of this project are given in references \cite{{Taqqu},{Pohl}}. The goal is to obtain a relative accuracy of $10^{-3}$. Then the effect of the proton size for the 1S level of hydrogen would be known with an uncertainty of $2.5~\mathrm{kHz}$. Using equation \ref{equ:Energie2}, it will be possible to deduce $R_{\infty}$ with a relative precision of about $2\times 10^{-12}$.

An alternative way to improve the precision of the Rydberg constant is to measure another optical frequency in hydrogen very precisely. Several groups are working in this direction. At the National Physical Laboratory, Flowers and colleagues have built a new experiment to improve the measurements of the $2\mathrm{S}-n\mathrm{S/D}$ hydrogen frequencies by using a femtosecond frequency comb \cite{Margolis}. In Paris, we intend to measure the optical frequency of the $1\rm S-3\rm S$ two-photon transition \cite{{Hagel},{Arnoult}}. This transition is also being studied by the group of H\"{a}nsch. The same group is also working towards an experiment on the $\mathrm{He}^+$ ion. All these efforts are promising to improve the accuracy on the Rydberg constant in the near future.
%\begin{table}
%\caption{Please write your table caption here.}
%\begin{tabular}{lll}
%\hline\noalign{\smallskip}
%\end{tabular}
%\end{table}
%

\end{document}